\documentclass[onecolumn]{aa}
\usepackage{graphicx}
\begin{document}
\def\simg{\mathrel{\hbox{\rlap{\lower.55ex \hbox {$\sim$}}
                   \kern-.3em \raise.4ex \hbox{$>$}}}}
\def\siml{\mathrel{\hbox{\rlap{\lower.55ex \hbox {$\sim$}}
                   \kern-.3em \raise.4ex \hbox{$<$}}}}
\title{On the difference between the short and long gamma-ray bursts}
\author{L.G. Bal\'azs \inst{1}
\and Z. Bagoly \inst{2} \and I. Horv\'ath \inst{3}
\and A.
M\'esz\'aros \inst{4} \and P. M\'esz\'aros \inst{5} }
\offprints{L.G. Bal\'azs} \institute{Konkoly Observatory,
Budapest, Box 67, H-1525, Hungary,
\email{balazs@ogyalla.konkoly.hu}
 \and Laboratory for Information
Technology, E\"{o}tv\"{o}s University,
  P\'azm\'any P\'eter s\'et\'any 1/A, H-1518,
Hungary, \email{bagoly@ludens.elte.hu}
 \and Dept. of Physics,
Bolyai Military University, Budapest, Box 12, H-1456, Hungary,
\email{hoi@bjkmf.hu}
 \and
Astronomical Institute of the Charles University, 180 00 Prague 8,
 V Hole\v sovi\v ck\'ach 2, Czech Republic,
\email{meszaros@mbox.cesnet.cz}
\and Dept. of Astronomy \&
Astrophysics, Pennsylvania State University, 525 Davey Lab,
\email{nnp@astro.psu.edu} }
\date{Received ...; accepted}
\authorrunning{Bal\'azs et al.}
\titlerunning{On the difference ... }

\abstract{ We argue that the distributions of both the intrinsic
fluence and the intrinsic duration of the $\gamma$-ray emission in
gamma-ray bursts from the BATSE sample are well represented by
log-normal distributions, in which the intrinsic dispersion is
much larger than the cosmological time dilatation and redshift
effects. We perform separate bivariate log-normal distribution
fits to the BATSE short and long burst samples. The bivariate
log-normal behaviour results in an ellipsoidal distribution, whose
major axis determines an overall statistical relation between the
fluence and the duration. We show that this fit provides evidence
for a power-law dependence between the fluence and the duration,
with a statistically significant different index for the long and
short groups. We discuss possible biases, which might affect this
result, and argue that the effect is probably real. This may
provide a potentially useful constraint for models of long and
short bursts.  }
 \maketitle
\section{Introduction}
\label{sec:intro} The simplest grouping of gamma-ray bursts
(GRBs), which is still lacking a clear physical interpretation, is
given by their well-known bimodal duration distribution. This
divides bursts into long ($T \simg 2$ s) and short ($T \siml 2$ s)
duration groups (\cite{k}), defined through some specific duration
definition such as $T_{90}$, $T_{50}$ or similar. The bursts
measured with the BATSE instrument on the Compton Gamma-Ray
Observatory are usually characterized by 9 observational
quantities, i.e. 2 durations, 4 fluences and 3 peak fluxes
(\cite{me96,pac,me01}). In a previous paper (\cite{bag98}) we used
the principal components analysis (PCA) technique to show that
these 9 quantities can be reduced to only two significant
independent variables, or principal components (PCs). These PCs
can be interpreted as principal vectors, which are made up of some
subset of the original observational quantities. The most
important PC is made up essentially by the durations and the
fluences, while the second, weaker PC is largely made up of the
peak fluxes. This simple observational fact, that the dominant
principal component consists mainly of the durations and the
fluences, may be of consequence for the physical modeling of the
burst mechanism. In this paper we investigate in greater depth the
nature of this principal component decomposition, and, in
particular, we analyze quantitatively the relationship between the
fluences and durations implied by the first PC.  In our previous
PCA treatment of the BATSE Catalog (\cite{pac}) we used
logarithmic variables, since these are useful for dealing with the
wide dynamic ranges involved. Since the logarithms of the
durations and the fluences can be explained by only one quantity
(the first PC), one might suspect the existence of only one
physical variable responsible for both of these observed
quantities. The PCA assumes a linear relationship between the
observed quantities and the PC variables. The fact that the
logarithmic durations and fluences can be adequately described by
only one PC implies a proportionality between them and,
consequently, a power law relation between the observed durations
and fluences. We analyze the distribution of the observed fluences
and durations of the long and the short bursts, and we present
arguments indicating that the intrinsic durations and fluences are
well represented by log-normal distributions. The implied
bivariate log-normal distribution represents an ellipsoid in these
two variables, whose major axis inclinations are statistically
different for the long and the short bursts. An analysis of the
possible biases and complications is made, leading to the
conclusion that the relationship between the durations and
fluences appears to be intrinsic, and may thus be related to the
physical properties of the sources themselves. We calculate the
exponent in the power-laws for the two types of bursts, and find
that for the short bursts the total emitted energy is weakly
coupled to the intrinsic duration, while for the long ones the
fluences are roughly proportional to the intrinsic durations. The
possible implications for GRB models are briefly discussed.

The paper is organized as follows.  Sect. 2 (Sect. 3) provides
classical $\chi^2$ fitting of the measured durations (fluences) -
separately for the short and long GRBs. The purposes of these two
Sections is to show that - separately for the two subgroups - both
the intrinsic durations and also the total emitted energies are
distributed log-normally. Using these results in Sect. 4 a
simultaneous fitting of the fluences and the measured durations
are done by the superposition of two bivariate log-normal
distributions. The purpose of this Section is to find the
power-law connections between the fluences and durations -
separately for the two subgroups. Because the observational biases
may play an essential role in these  results the biases are
studied in this Section, too. Sect. 5 discusses and summarizes the
results of article. In the Appendix some technical calculations
are presented.

\section{Analysis of the duration distribution}
\label{sec:durations}

Our GRB sample is selected from the Current BATSE Gamma-Ray Burst
Catalog according to two criteria, namely, that they have both
measured $T_{90}$ durations and fluences (for the definition of
these quantities see \cite{me01}, henceforth referred to as the
Catalog). The Catalog in its final version lists 2041 bursts for
which a value of $T_{90}$ is given. The fluences are given in four
different energy channels, $F_1, F_2, F_3, F_4$, whose energy
bands correspond to $[25,50]$ keV, $[50,100]$ keV, $[100,300]$ keV
and $> 300 $ keV. The "total" fluence is defined as
$F_\mathrm{tot} = F_1 + F_2 + F_3 + F_4$. We restrict our sample
to include only those GRBs, which have $F_i>0$ values in the first
three channels; i.e. $F_1, F_2, F_3$ are given. Concerning the
fourth channel, whose energy band is $>300$ keV, if we had
required $F_4>0$ as well, this would have reduced the number of
eligible GRBs by $\simeq 20\%$. Hence, we decided to accept also
these bursts with $F_4 = 0$, rather than deleting them from the
sample. (With this choice we also keep in the sample the
no-high-energy (NHE) subgroup defined by \cite{pen97}.) Our choice
of $F\equiv F_\mathrm{tot}$, instead of some other quantity as the
main variable, is motivated by two arguments. First, as discussed
in \cite{bag98}, $F_\mathrm{tot}$ is the main constituent of one
of the two PCs which represent the data embodied in the BATSE
Catalog, and hence it can be considered as a primary quantity,
rather than some other combination or subset of its constituents.
Second, Petrosian and collaborators in a series of articles
(\cite{ep92}, \cite{pl96}, \cite{lp96}, \cite{lp97}) have also
argued for the use of the fluence as the primary quantity instead
of, e.g., the peak flux. Using such defined $F_\mathrm{tot}$, from
the sample only such GRBs are deleted, which have no measured
$F_\mathrm{tot}$. Because also the peak fluxes are needed, too, we
are left with $N=1929$ GRBs, all of which have defined $T_{90}$
and $F_\mathrm{tot}$, as well as peak fluxes $P_{256}$ on the 256
ms trigger scale. If the peak flux $P_{64}$ on the 64 ms trigger
scales is needed, then the sample contains $N=1972$ GRBs. These
are the samples studied in this article.

The distribution of the logarithm of the observed $T_{90}$
displays two prominent peaks\footnote{There is also an evidence
for the existence of a third intermediate subgroup as part of the
long duration group (\cite{hor}, \cite{muk}, \cite{hak00a},
\cite{HA}, \cite{brc}, \cite{hor2}), which shows a distinct sky
angular distribution (\cite{mesz00a}, \cite{mesz00b},
\cite{li01}). We do not deal with this third group here.}, which
is interpreted as reflecting the existence of two groups of GRBs
(\cite{k}, \cite{no00}). This bimodal distribution can be well
fitted by the sum of two Gaussian distributions (\cite{hor})
indicating that both the long and the short bursts are
individually well fitted by pure Gaussian distributions in the
logarithmic durations. The fact that the distribution of the BATSE
$T_{90}$ quantities within a group is log-normal is of interest,
since we can show that this property may be extended to the {\it
intrinsic} durations as well. Let us denote the observed duration
of a GRB with $T_{90}$ (which may be subject to cosmological time
dilatation), and denote with $t_{90}$ the duration which would be
measured by a comoving observer, i.e. the intrinsic duration. One
has
\begin{equation}
 T_{90} = t_{90} f(z),
\label{eq:t90}
\end{equation}
where $z$ is the redshift, and $f(z)$ measures the time
dilatation. For the concrete form of $f(z)$ one can take $f(z) =
(1+z)^k$, where $k=1$ or $k=0.6$, depending on whether energy
stretching is included or not (\cite{fb95}, \cite{mm96}). If
energy stretching is included, for different photon frequencies
$\nu$ the $t_{90}$ depends on these frequencies as $t_{90}(\nu) =
t_{90}(\nu_\mathrm{o}) (\nu/\nu_\mathrm{o})^{-0.4} \propto
\nu^{-0.4}$, where $\nu_\mathrm{o}$ is an arbitrary frequency in
the measured range (i.e. for higher frequencies the intrinsic
duration is shorter). The observed duration at $\nu$ is simply
$(1+z)$ times the intrinsic duration at $\nu \times (1+z)$. Thus,
$T_{90}(\nu) = t_{90}(\nu (1+z)) (1+z)$ =
$t_{90}(\nu_\mathrm{o})(\nu (1+z)/\nu_\mathrm{o})^{-0.4} (1+z) =
t_{90}(\nu)(1+z)^{0.6}$. Hence, when stretching is included, $f(z)
= (1+z)^{0.6}$ is used. Taking the logarithms of both sides of
Eq.~(1) one obtains the logarithmic duration as a sum of two
independent stochastic variables. According to a mathematical
theorem of Cram\'er (\cite{cra37}, \cite{renyi}), if a variable
$\zeta$ - which has a Gaussian distribution - is given by the sum
of two independent variables, e.g. $\zeta = \xi + \eta$, then both
$\xi$ and $\eta$ have Gaussian distributions. (In practical cases,
however, this holds, of course, only if the variances of $\xi$ and
$\eta$ are comparable. If the variance of, say, $\xi$ is much
smaller than the variance of $\eta$, then both the variables
$\zeta$ and $\eta$ may have a normal distribution - but nothing
can be said about the distribution of $\xi$. It can, but also need
not be Gaussian.)

Therefore, the Gaussian distribution of
$\log T_{90}$ - confirmed for the long and short groups separately
(\cite{hor}) - implies that the same type of distribution exists for
the variables $\log t_{90}$ and $\log f(z)$. However, unless the
space-time geometry has a very particular structure, the
distribution of $\log f(z)$ cannot be Gaussian. This means that
the Gaussian nature of the distribution of $\log T_{90}$ must be
dominated by the distribution of $\log t_{90}$ {\it alone}, and therefore the
latter must then necessarily have a Gaussian distribution. In other words,
the variance of $f(z)$ must be much smaller than the variance of
$\log t_{90}$. This must
hold for both duration groups separately. This also implies that
the cosmological time dilatation should not affect significantly
the observed distribution of $T_{90}$, which therefore is not
expected to differ statistically from that of $t_{90}$. We note
that several other authors (\cite{wp94}, \cite{nor94},
\cite{nor95}) have already suggested that the distribution of
$T_{90}$ reflects predominantly the distribution of $t_{90}$. Nevertheless,
our argumentation based on the mathematical Cram\'er theorem is new.

One can check the above statement quantitatively by calculating
the standard deviation of $f(z)$, using the available observed
redshifts of GRB optical afterglows. The number of the latter is,
however, relatively modest, and, in addition,  so far they have
been obtained only for long bursts. There are currently upwards of
21 GRBs with well-known redshifts (\cite{gre02}).
 The calculated standard deviation is $\sigma_{\log f(z)}=0.17$,
assuming $\log f(z)=\log (1+z)$.  Comparing the variance
$\sigma^2_{\log f(z)}$ with that of the group of long burst
durations which gives $\sigma_{\log T_{90}} = 0.5$, one infers
that the variance of $\log f(z)$, or of $\log (1+z)$, can explain
maximally only about $(0.17/0.50)^2 \simeq 12\%$ of the total
variance of the logarithmic durations. (If $ f(z) = (1+z)^{0.6}$,
then the variance of $\log f(z)$ can only explain an even smaller
amount, because $\sigma_{\log f(z)} = 0.6\times 0.17$.) This
comparison supports the conclusion obtained by applying Cram\'er's
theorem to the long duration group. For the short duration group,
since this does not so far have measured redshifts, one can rely
only on the theorem itself.
\section{Analysis of the fluence distribution}
\label{sec:energies}

The observed total fluence $F_\mathrm{tot}$ can be expressed as
\begin{equation}
F_\mathrm{tot} = \frac{(1+z)E_\mathrm{tot}}{4\pi d_l^2(z)} = c(z)
E_\mathrm{tot}. \label{eq:fluence}
\end{equation}
Here $E_\mathrm{tot}$ is the total emitted energy of the GRB at
the source in ergs, the total fluence has dimension of erg/cm$^2$,
and $d_\mathrm{l}(z)$ is the luminosity distance corresponding to
$z$ for which analytical expressions exist in any given Friedmann
model (\cite{weinberg72}, \cite{peebles93}). (We note that the
considerations in this paper are valid for any Friedmann  model.
Note also that the usual relation between the luminosity and flux
is given by a similar equation without the extra $(1+z)$ term in
the numerator. Here this extra term is needed because both the
left-hand-side is integrated over the observer-frame time and the
right-hand-side is integrated over the time at the source
(\cite{mm95}).)

Assuming as the null hypothesis that the $\log F_\mathrm{tot}$ of
the short bursts has a Gaussian distribution, for the sample of
447 bursts with $T_{90} < 2$ s, a $\chi^2$ test with 26 degrees of
freedom gives an excellent fit with $\chi^2=20.17$. Accepting the
hypothesis of a Gaussian distribution within this group, one can
apply again Cramer's theorem similarly to what was done for the
logarithm of durations. This leads to the conclusion that either
{\it both} the distribution of $\log c(z)$ and the distribution of
$\log E_\mathrm{tot}$ are Gaussian ones, or else the variance of
one of these quantities is negligible compared to the other, which
then must be mainly responsible for the Gaussian behaviour.
Because $\log c(z)$ hardly can have a log-normal distribution, the
second possibility seems to be the situation. In any case, one may
conclude that the intrinsic fluence (i.e. the total emitted
energy) should be distributed log-normally.

In the case of the long bursts, a fit to a Gaussian distribution
of logarithmic fluences does not give a significance level, which
is as convincing as for the short duration group. For the 1482
GRBs with $T_{90} > 2$ s a $\chi^2$ test on $\log F_\mathrm{tot}$
with 22 degrees of freedom gives a fit with $\chi^2=35.12$.
Therefore, in this case the $\chi^2$ test casts some doubt on
normality but only with a  relatively high error probability of
3.5\% for rejecting a Gaussian distribution (\cite{tw53},
\cite{KS76}, \cite{pre}). This circumstance prevent us from
applying Cram\'er's theorem directly in the same way as we did
with the short duration group. Calculating the variance of $\log
c(z)$ for the GRBs with known 21 redshifts (\cite{gre02}) one
obtains $\sigma_{\log c(z)}=0.43$. For the GRBs of long duration,
however, one obtains $\sigma_{\log F_\mathrm{tot}} = 0.66$. The
ratio of these variances  equals $(0.43^2/0.66)^2 \simeq 42 \%$,
i.e. more than  half of the variance of $F_\mathrm{tot}$ is not
explained by the variance of $c(z)$. (If one takes into account
the energy stretching even a larger fraction remains unexplained).
In other words, a significant fraction of the total variance of
$F_\mathrm{tot}$ has to be intrinsic. It is worth mentioning that
the unexplained part of the variance of $F_\mathrm{tot}$
corresponds nicely to the value obtained in Sect.
\ref{sec:correlation} making use the EM algorithm.

Despite these difficulties, there is a substantial reason to argue
that the intrinsic distribution of total emitted energies is
distributed log-normally for the long subgroup, too.

The Gaussian behaviour of $\log c(z)$ can almost certainly be
excluded. One can do this on the basis of the current observed
distribution of redshifts (\cite{gre02}), or on the basis of fits
of the number vs. peak flux distributions (\cite{fb95},
\cite{uw95}, \cite{hor3}, \cite{rm97}). In such fits, using a
number density $n(z) \propto (1+z)^D$ with $D \simeq (3-5)$, one
finds no evidence for the stopping of this increase with
increasing $z$ (up to $z \simeq (5-20)$). Hence, it would be
contrived to deduce from this result that the distribution of
$\log c(z)$ is normal. In order to do this, one would need several
ad hoc assumptions. First, the increasing of number density would
need to stop around some unknown high $z$. This was studied
(\cite{mm95}, \cite{hor3}, \cite{mm96}), and no such effect was
found. (For the sake of preciseness it must be added here that
these fits were done for the whole sample of GRBs. But, because
GRBs are dominated by the long ones, conclusions from these fits
should hold for the long subgroup, too.)
 Second, even if this were the case, above this $z$ the
decrease of $n(z)$ should mimic the behavior of a log-normal
distribution for $c(z)$, without any obvious justification. Third,
below this $z$ one must again have a log-normal behavior for
$c(z)$, in contradiction with the various number vs. peak flux
fits. Fourth, this behavior should occur for any subclass
separately. Hence, the assumption of log-normal distribution of
$c(z)$ appears highly improbable.

Having a highly improbable log-normal distribution of $\log c(z)$,
which variance is surely not negligible, a 3.5\% error probability
(i.e the probability that we reject the hypothesis of normality
but it is still true) from the goodness-of-fit is still
remarkable. One may argue that, if the distribution of the total
emitted energy were not distributed log-normally, then the two
non-normal distributions together would give a fully wrong
$\chi^2$ fit for $\log F_\mathrm{tot}$; under this condition even
the 3.5\% probability would not be reachable. Of course, this
argumentation is more or less heuristic, and - as the conclusion -
one cannot say that the log-normal distribution of
$E_\mathrm{tot}$ is confirmed similarly unambiguously in both
subgroups. In the case of long subgroup questions still remain,
and they will still be discussed (end of Sect. 4.3). It is worth
mentioning the rise times, fall times, FWHM, pulse amplitudes and
areas were measured and the frequency distributions are
consistent with log-normal distributions (\cite{mcb01},
\cite{qui02}).

In addition, even in the case of short GRBs the situation is not
so clear yet. The argument based on the Cram\'er theorem for the
short GRBs should also be taken with some caution. As shown in
\cite{bag98}, the stochastic variable corresponding to the
duration is independent from that of the peak flux. This means
that a fixed level of detection, given by the peak fluxes, does
not have significant influence on the shape of the detected
distribution of the durations (\cite{ep92}, \cite{wp94},
\cite{nor94}, \cite{nor95}, \cite{pl96}, \cite{lp96},
\cite{lp97}). In the case of the fluences, however, a detection
threshold in the peak fluxes induces a bias on the true
distribution, since fluences and durations are stochastically not
independent. Therefore, the log-normal distribution recognized
from the data does not necessarily imply the same behaviour for
the true distribution of fluences occurring at the detector. In
other words, observational biases may have important roles; in
addition, for both subgroups. A discussion of these problems can
be found in a series of papers published by Petrosian and
collaborators (\cite{ep92}, \cite{pl96}, \cite{lp96}, \cite{lp97},
\cite{lp99}). In what follows, we also will study the biases
together with the fitting procedures.

\section{Correlation between the fluence and duration}
\label{sec:correlation}

In the previous Sections we presented firm evidences that the
observed distribution of the durations is basically intrinsic. We
argued furthermore that a significant fraction of the variance of
the fluences is also intrinsic. We proceed a step further in this
Section and try to demonstrate that there is a relationship
between the duration and the fluence which is also intrinsic.
There are two basic difficulties in searching the concrete form of
this relationship (if there is any at all): first, we observe only
those bursts which fulfill some triggering criteria and, second,
the observed quantities are suffering from some type of bias
depending on the process of detection. Several papers discuss
these biases (\cite{ep92}, \cite{lamb93}, \cite{lp96},
\cite{pl96}, \cite{lp97}, \cite{stern}, \cite{pac}, \cite{hak00},
\cite{meg00}).
 In the following we will address these issues in a new way.

The detection proceeds on three time scales: the input signal is
analyzed on 64, 256 and 1024 ms resolution. The counts in these
bins of these scales are compared with the corresponding 17 second
long averaged value. There are eight detectors around the BATSE
instrument. If at least one of the three peak intensities in the
second brightest detector exceeds 5.5 sigma of the threshold
computed from the averaged signal the burst will be detected. In
case of the bursts of long duration (at least several seconds) the
differences in the time scales of detection do not play an
important role since the vast majority of the events were
triggered on the 1024 ms scale and the detection proceeded if the
peak exceeded the threshold on this time scale. In contrast, at
the bursts of short duration - when $T_{90}$ could be much shorter
then the time scale of the detection - the situation could be
drastically changed. Looking at the data of the BATSE the bursts
of duration of $T_{90}<2 s$ are mixtures of those triggered on
different time scales. Among bursts triggered on the same time
scale the detection proceeds when the corresponding peak flux
exceeds the threshold. In the case of bursts, which are shorter
than their triggering time scale, the corresponding peak fluxes
are given by the fluence itself. This has the consequence that the
threshold in the peak flux means the same for the fluence, i.e. it
results a horizontal cut on the fluence - duration plane and a
bias in the relationship between these quantities. In order to
minimize this effect we will use the peak flux on the 64 ms and
256 ms time scale in our further analysis. The BATSE had a
spectral response on the detected  $ \gamma $ radiation. It had
the consequence that different measured values were assigned to
bursts having the same total energy at the entrance of the
detector if the incoming photons had different spectral
distributions.

The duration of a GRB is only a lower
limit for its  intrinsic value since a certain fraction of the
burst can be buried in the background noise. Therefore any
relationship recognized among the observed fluence and duration is
not necessarily representative for those between the corresponding
intrinsic quantities. In the next paragraphs  we address these
issues in more details.

\subsection{Effect of the detection threshold on the joint probability
distribution of the fluence and duration}

In the following we will study the effect of the detection
threshold on the joint probability distribution of the observed
fluence and duration. In order to put this effect into a
quantitative basis we use the law of full probabilities (see e.g.
\cite{renyi}). Let $P(F_\mathrm{tot},T_{90})$ be the joint
probability density of the fluence and duration. Using this
theorem any of the probability densities on the right side can be
written in the form of
\begin{equation}
 P(F_\mathrm{tot}, T_{90}) = \int \limits_0^{\infty}P(F_\mathrm{tot}, T_{90}|p) G(p) dp,
\end{equation}
where $p$ is the peak flux at any of the 64 ms, 256 ms and 1024 ms
time scales, $P(F_\mathrm{tot}, T_{90}|p)$ is the joint
(bivariate) probability density of the fluence and duration
(assuming that $p$ is given), and $G(p)$ is the probability
density of $p$. This means that, if there are $N$ bursts in the
sample, then $N P(F_\mathrm{tot},T_{90}) d \log F_\mathrm{tot} d
\log T_{90}$ is the expected number of observed GRBs in the
infinitesimal intervals $[\log F_\mathrm{tot}, (\log
F_\mathrm{tot} + d \log F_\mathrm{tot})]$ and $[\log T_{90}, (\log
T_{90} + d \log T_{90})]$, respectively. Among the bursts
triggered on a given time scale $G(p)$ represents an unbiased
function above $p_\mathrm{th}$, the peak flux corresponding to the
detection threshold. Below this limit, however, $G(p)$ is biased
by the process of detection. It inserts also a bias on the joint
probability density of the observed fluence and duration.
Nevertheless, the kernel $P(F_\mathrm{tot}, T_{90}|p)$ represents
some intrinsic relationship between these two quantities, and it
is free from the bias of $G(p)$. Following our discussion given
above, we use in the following the peak fluxes of the 64 ms time
scale.
\subsection{Intrinsic relationship between the fluence and the duration}
We demonstrated in an earlier paper (\cite{bag98}) that the
logarithms of the peak flux and the duration represent two
independent stochastic variables and the logarithmic fluence can
be well approximated as the linear combination of these variables:
\begin{equation}
\log F_\mathrm{tot} = a_1 \log T_{90} + a_2 \log p + \epsilon,
\end{equation}
where $a_1$, $a_2$ are constants, and $\epsilon$ is a noise term (later on we
will see that $a_1$ may depend on the duration, i.e. it is different for the
bursts of short and long duration). One may confirm
this statement by inspecting
the Tables given in the Appendix. They demonstrates convincingly that,
independently of the choice of the peak flux, the standard deviation and the
mean value of the duration is not changed significantly.
This expression reveals that - fixing the peak intensity - the distribution of
the fluences reflects basically the distribution of the durations. Since the
probability density of the durations is a superposition of two Gaussian
distributions, the
same should hold also for the fluences. Consequently, we may assume that the
joint conditional probability distribution of the fluence and duration
consists of a superposition of two two-dimensional Gaussian distributions.
One such distribution takes the form
\begin{equation}
 f(x,y) dx dy =
\frac{N}{2 \pi \sigma_x \sigma_y \sqrt{1-r^2}}\times
\exp\left[-\frac{1}{2(1-r^2)}\left(\frac{(x-a_x)^2}{\sigma_x^2} +
\frac{(y-a_y)^2}{\sigma_y^2} - \frac{2r(x-a_x)(y-a_y)}
{\sigma_x \sigma_y}\right)\right]
 dxdy\;,
\label{eq:exp}
\end{equation}
where $x =\log T_{90}$, $y = \log F_\mathrm{tot}$ $a_x$, $a_y$ are
the means, $\sigma_x$, $\sigma_y$ are the dispersions, and $r$ is
the correlation coefficient (\cite{tw53}; Chapt. 1.25). In our
case one needs a weighted sum of two such bivariate distributions.
This means that 11 free parameters should be determined (two times
5 parameters for the both distributions; the 11th independent
parameter is the weight of the, say, first subgroup). This also
means that two $r$ correlation coefficients should be obtained,
which may be different for the two subgroups.

The parameters $a_x, \sigma_x$, characterizing the distribution of
the duration do not depend on the peak flux, because $T_{90}$ and
$p$ are independent stochastic variables. The only dependent
parameter is $a_y$, the mean value of the fluence. In the case,
when the $r$-correlation coefficient differs from zero, the
semi-major axis of the dispersion ellipse represents a linear
relationship between $\log T_{90}$ and $\log F_\mathrm{tot}$, with
a slope of $m=\tan \alpha$, where
\begin{equation}
\tan 2 \alpha = \frac{2r\sigma_x \sigma_y}{\sigma_x^2 - \sigma_y^2}\;.
\label{eq:tan}
\end{equation}
This linear relationship between the logarithmic variables implies
a power-law relation of form $F_\mathrm{tot} = (T_{90})^m$ between
the fluence and the duration, where $m$ may be different for the
two groups. Replacing the $G(p)$ probability density by the
empirical distribution of the measured peak fluxes, one may write
the joint probability density of the fluence and duration in the
form of
\begin{equation}
 P(F_\mathrm{tot}, T_{90}) = \int \limits_0^{\infty}P(F_\mathrm{tot}, T_{90}|p) G(p) dp
\simeq \sum \limits_{i=1}^N P(F_\mathrm{tot}, T_{90}|p_i) \simeq
\sum \limits_{l=1}^k b_l P(F_\mathrm{tot}, T_{90}|p_l),
\end{equation}
i.e. the integral is approximated by a sum of $k$ separate terms (bins), in
which $b_l$ is the number of GRBs at the given bin.

The $k$ is the number of bins at the right-hand-side, and is somewhat
arbitrary. Trivially, bigger $k$ leads to a better approximation of
the integral. On the other hand, bigger $k$ leads to the situation, when in
one single bin the number of GRBs $b_l$ is smaller. Hence,
$k$ should be small in order
to get enough number of GRBs in each bin for making statistics, but not
too small in order to have good approximation of the integral.

\subsection{Maximum Likelihood estimation of the parameters via EM algorithm}
One finds in the Tables of Appendix the computed mean values and
standard deviations of the logarithmic durations for the short and
long bursts, respectively. These Tables clearly suggest that,
except for the faintest bins where we expect serious biases in the
duration and fluence due to the detection close to the background,
the standard deviations do not differ significantly between the
bins. Dividing the sample into short and long bursts by the cut of
$T_{90}<2s$ and $T_{90}>2s$, we may assume that these subsamples
are dominated by only one Gaussian distribution and we may compute
its parameters in a simple way as given below. If the
$P(F_\mathrm{tot}, T_{90}|p)$ conditional probability density is a
pure Gaussian one, then the Maximum Likelihood ($ML$) estimation
of its parameters would be very simple, because they can be
obtained by computing the mean values, standard deviations and the
correlation  between the fluence and duration. In the reality,
however, this probability density is a superposition of two
Gaussians one, and the simple cut at $T_{90}=2s$ is hardly
satisfactory. The proper way to estimate the free parameters is
not so simple. For this reason we will use a procedure, called EM
algorithm ({\bf E}xpectation and {\bf M}aximalization), which
terminates at the $ML$ solution (\cite{demp77}). If we knew, which
of the bursts belong to the short and long duration groups, we may
add a $\{i_1,i_2\}$ two dimensional indicator variable to each GRB
having the value of $\{1,0\}$ in the case if a burst was short,
and similarly $\{0,1\}$ if it was long. The sample means of
$T_{90}$ weighted with $i_1$ would give the $ML$ estimation of
$a_x$ of the first Gaussian distribution (i.e. $ a_x = \sum
\nolimits_{j=1}^N i_{1j}x_j/\sum \nolimits_{j=1}^n i_{1j})$. The
same hold for the other parameters. Weighting with $i_2$ would
give the parameters of the second Gaussian distribution. Hence 10
parameters of the two distributions would be well calculable. The
11th parameter would also be trivially calculable, because the
fraction of first subgroup should simply be $\sum_{i=1}^{N}
i_1/N$. Hence, if the values of the  $\{i_1,i_2\}$ indicator
variable were known,  the $ML$ parameters would be well
calculable.

If the parameters of the two Gaussians were given, one could
compute the $\left\{p_1,p_2\right\}$  membership probabilities of
a burst to each of the two groups. Replacing the indicator
variable by these probabilities one may calculate new parameters
in the same way as was done assuming $\left\{i_1,i_2\right\}$ were
given. Then one may again calculate new $\left\{i_1,i_2\right\}$, and
 again the new parameters. This iteration
is exactly the procedure, what EM algorithm is doing. One gives an
initial estimate for the parameters of the two Gaussian
distributions. Then one estimates the membership probabilities (E
step). Weighting with the membership probabilities one obtains the
new $ML$ estimation of the parameters (M step). Repeating these
steps successively one proceeds to the $ML$ solution of the
parameter estimation (\cite{demp77}).

In order to fit the $[\log T_{90}, \log F_\mathrm{tot}]$ data
pairs with the superposition of two two-dimensional Gaussian
bivariate distributions we splitted the Catalog into subsamples
with respect to 64 ms peak fluxes. The strata were obtained by
taking 0.2 wide strips in the logarithmic peak fluxes. Table 1.
summarizes the number of GRBs within the strata. In addition, also
the number of GRBs with $T_{90} < 2 s$ and with $T_{90} < 0.064 s$
are given there. The first one shows that, roughly, which fraction
of GRBs belong to the short subgroup in the given strata, and the
second one shows which fraction is maximally biased.

\begin{table}
\caption{ Number of GRBs within the 0.2 wide strata of the
logarithmic 64 ms peak fluxes. \newline}
\begin{tabular}{|c|cccc|}
\hline Serial No. & $ \log P_{64}$ & total No.
& No. of GRBs & No. of GRBs    \\
       &    &  of GRBs & with $T_{90} < 2 s$ & with $T_{90} < 0.064 s$ \\
 \hline
       \  \  1.  &    -0.6  -  -0.4  & 5 & 1 & 0         \\
       \   \ 2.  &     -0.4  -  -0.2   &  113 & 5 & 0 \\
         \ \ 3.  &   -0.2  - 0.0   &  385 & 44 & 1 \\
         \ \ 4.  &    0.0 - 0.2    &    434 & 104 & 4 \\
         \ \ 5.  &    0.2 - 0.4    &   365 & 126  & 8 \\
         \ \ 6.  &   0.4  -  0.6   &   254 & 79 & 3 \\
         \ \ 7.  &   0.6  - 0.8    &    166 &47 & 2 \\
         \ \ 8.  &    0.8 - 1.0    &     95 & 34 & 0 \\
        \ \ 9.  &     1.0 - 1.2    &     74 & 22 & 0 \\
       \ 10.  &      1.2 - 1.4    &       39 & 6 & 0\\
       \ 11.  &    1.4 - 1.6     &      19 & 5 & 0 \\
       \ 12.  &    1.6 - 1.8     &     15 & 2 &  0\\
       \ 13.  &    1.8 - 2.0     &      6 & 1 &  0\\
       \ 14.  &     2.0  $<$      &     2 & 0 & 0 \\
\hline
\end{tabular}
\end{table}

In the fitting procedure we omitted bins No. 1.--3., being
affected by selection bias, and also No. 9.--14., being scarcely
populated. We performed the $ML$ fitting in the bins No. 4.--8.,
making use the EM algorithm. Table 2. summarizes the results of
the $ML$ fitting for the short GRBs, and Table 3 for the long
GRBs, respectively. On Figs. 1-5 the results of fitting for bins
No. 4.--8. are shown. The ellipses define the 1-sigma and 2-sigma
regions, respectively.

\begin{table}
\caption{Results of the $ML$ fitting for the short GRBs  using the
EM algorithm. \newline Weighted mean for $m$ is $m =0.81 \pm
0.06$.
\newline}
\begin{tabular}{|c|cccccc|c|c|}
\hline Strip No.  & rel. frequency  & $a_x$ & $a_y$  & $\sigma_x$
& $\sigma_y$  & $r$ & total No. of GRBs & $m=tan \alpha$ \\ \hline
 4.  &   .293  &   -.199 &   -6.587  &   .549 &
 .502 &    .593  &    434  &  0.86 \\
  5.    &  .418   &  -.275  &   -6.488  &
    .575 &    .503 &    .591  &   365  &  0.80 \\
 6.    &  .321   &  -.365  &  -6.244   &  .486  &
  .497  &   .515   &  254  & 1.04 \\
 7.    &  .332   &  -.188  & \   -5.921
  &  .510  &   .420  &   .342    &  166 & 0.58 \\
 8.     &  .358   &  -.325  &  \ -5.910   &  .440  &
   .347   &  .279    & \   95  &  0.46\\
\hline
\end{tabular}
\end{table}
\begin{figure*}
\centering
\includegraphics{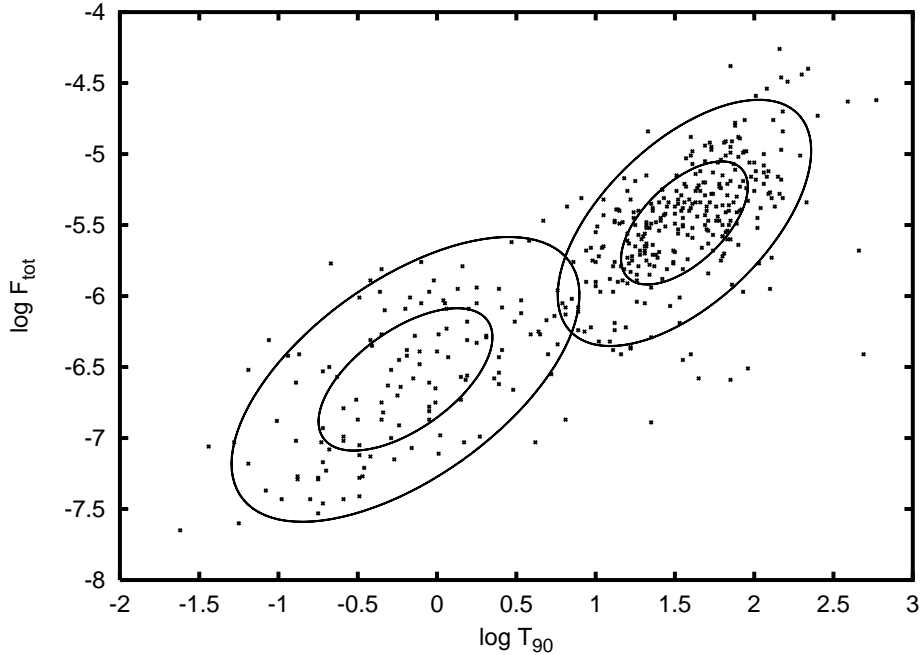}
\caption{The best $ML$ fits of the two log-Gaussian distributions
for the faintest sample No.4 with $N=434$.}
\end{figure*}

\begin{figure*}
\centering
\includegraphics{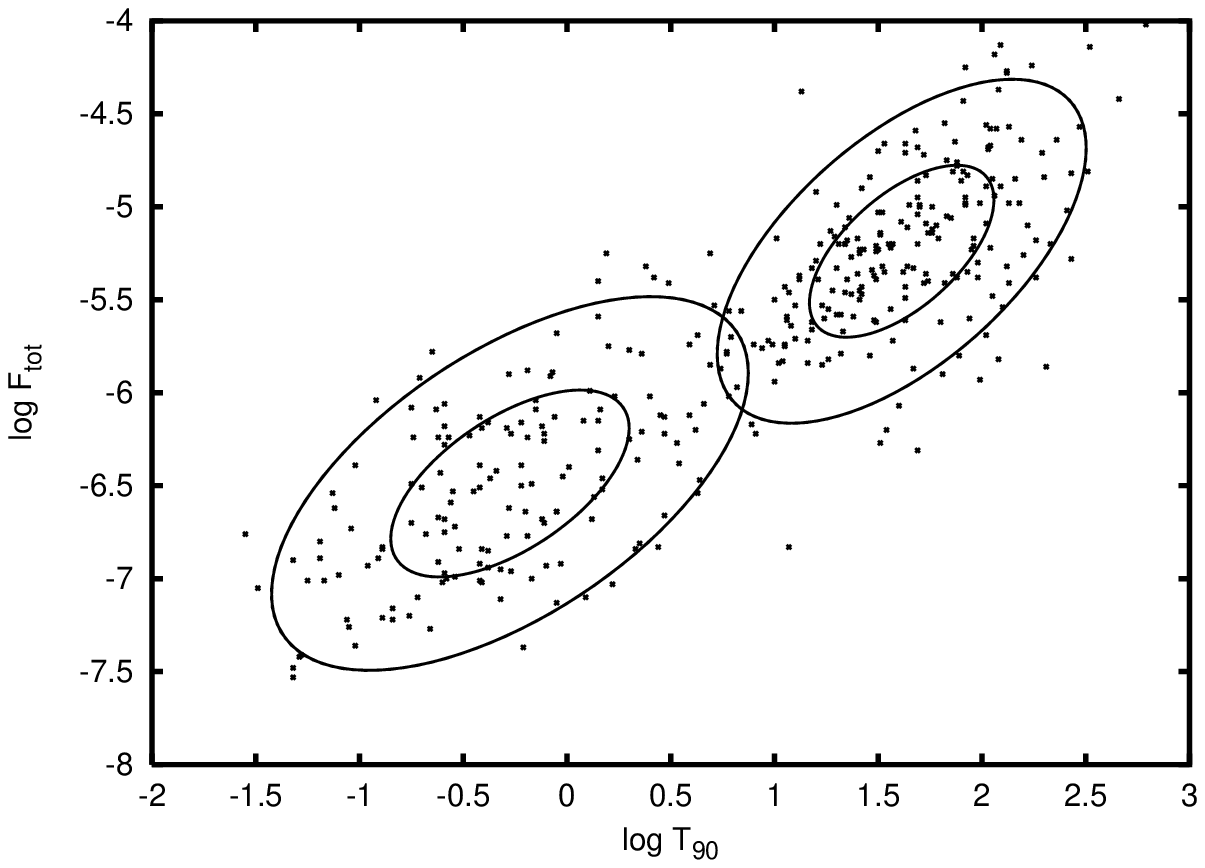}
\caption{The best $ML$ fits of the two log-Gaussian distributions
for the sample No.5 with $N=365$.}
\end{figure*}

\begin{figure*}
\centering
\includegraphics{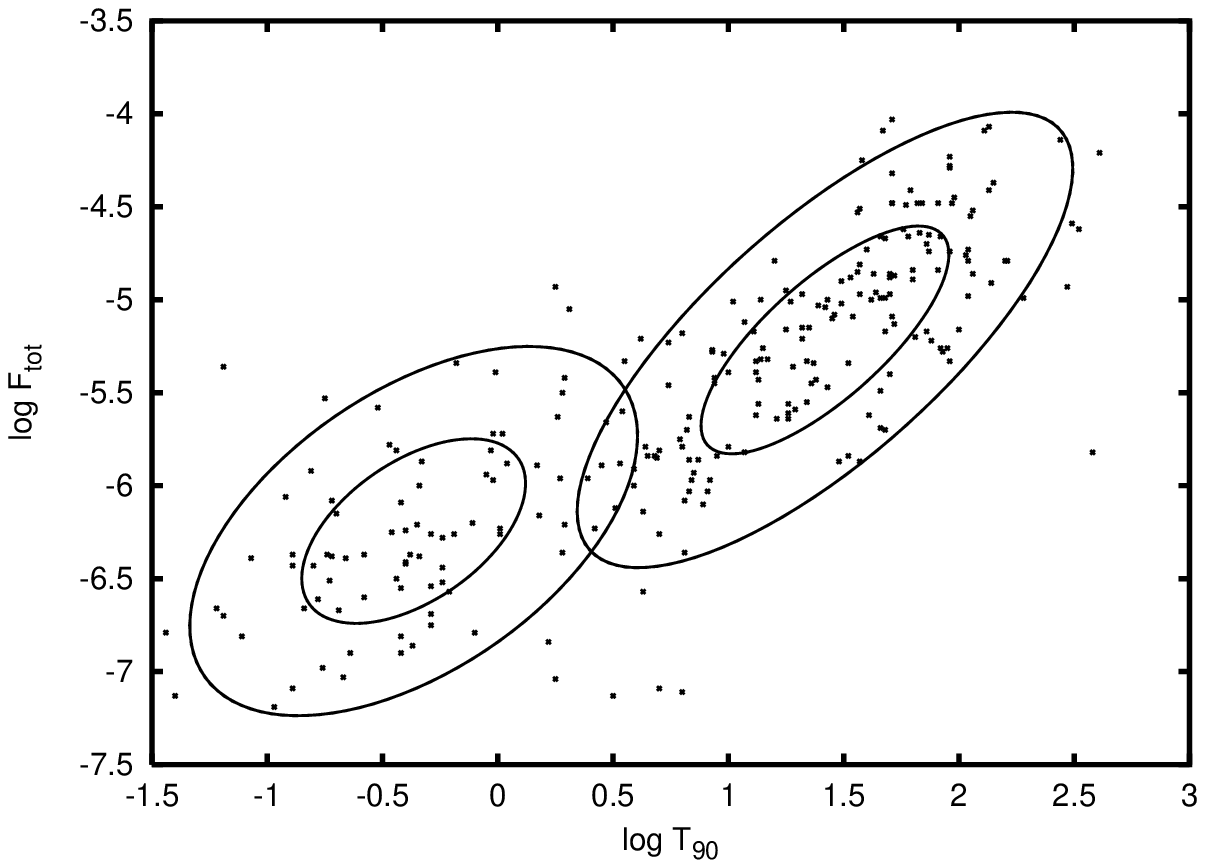}
\caption{The best $ML$ fits of the two log-Gaussian distributions
for the sample No.6 with $N=254$.}
\end{figure*}

\begin{figure*}
\centering
\includegraphics{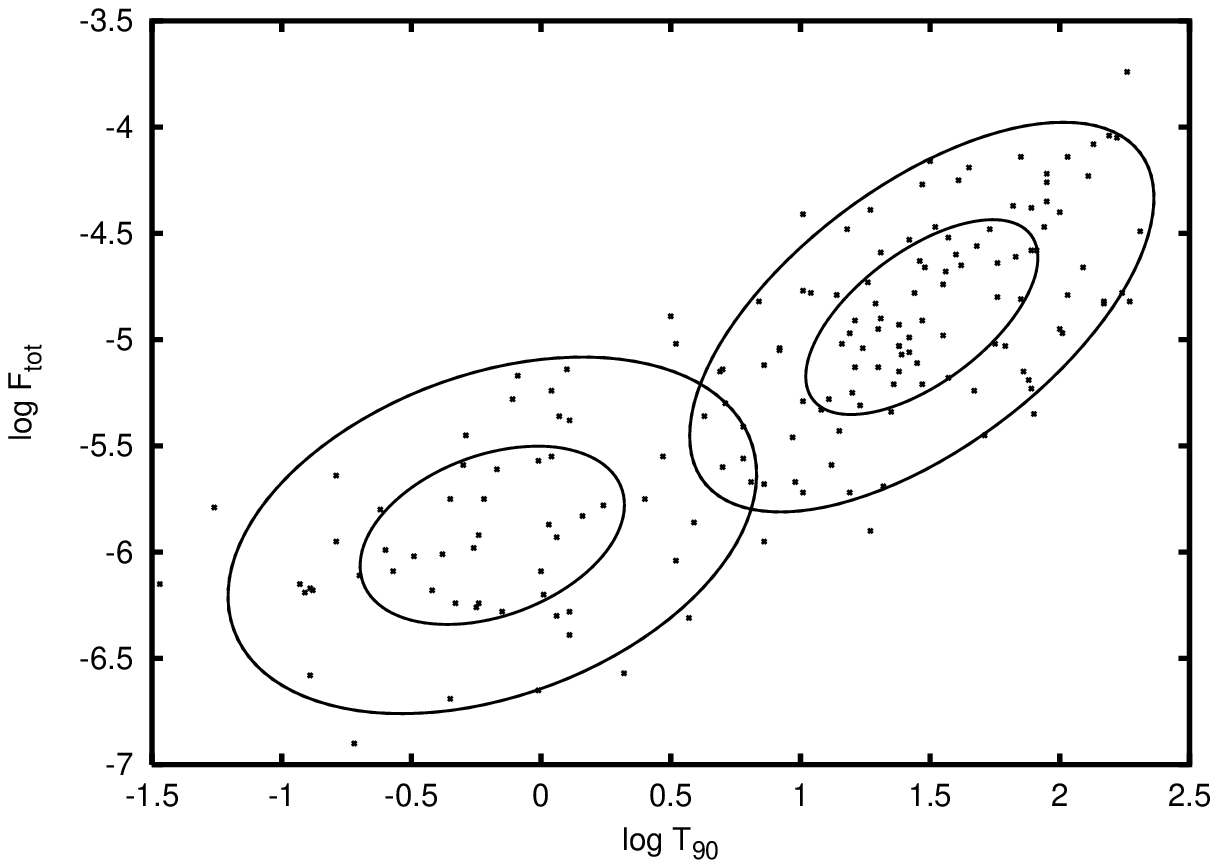}
\caption{The best $ML$ fits of the two log-Gaussian distributions
for the sample No.7 with $N=166$.}
\end{figure*}

\begin{figure*}
\centering
\includegraphics{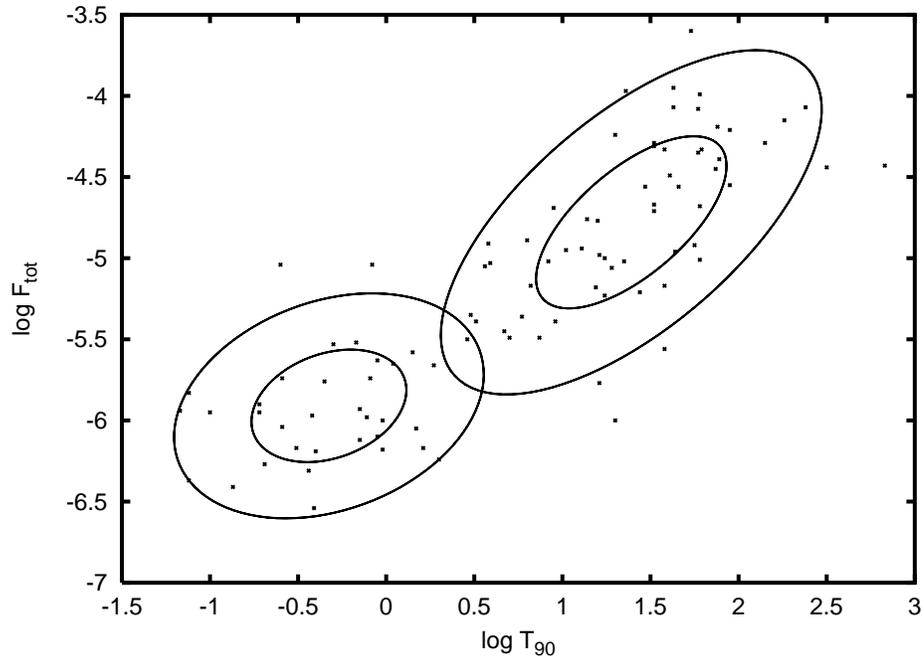}
\caption{The best $ML$ fits of the two log-Gaussian distributions
for the brightest sample No.8 with $N=95$.}
\end{figure*}

The slopes of short GRBs obtained for the bins No.7. and  No.8.
differ remarkably from those in bins No. 4.--6. They are based,
however, on a small number of bursts; hence, the $r$ parameter is highly
uncertain. Since we used for weighting the number of GRBs within
the given bin their, contribution to the final result is marginal.
\begin{table}
\caption{Results of the $ML$ fitting for the long GRBs  using the
EM algorithm. \newline Weighted mean for $m$ is $m =1.11 \pm
0.03$.
  \newline}
\begin{tabular}{|c|cccccc|c|c|}
\hline Strip No.  & rel. frequency  & $a_x$ & $a_y$  & $\sigma_x$
& $\sigma_y$  & $r$ & total No. of GRBs & $m=tan \alpha$ \\ \hline
  4.   &    .707  &   1.560  &  -5.485    &
  .400   &  .434   &  .586  &  434    &  1.15\\
5.   &  .582   & 1.613    & -5.239
 &   .445   &  .463   &  .599  &    365  &   1.07 \\
   6.  &     .679 &   1.419  &   -5.216    & .538
     & .613    & .753    &  254   & 1.19 \\
7.  &     .668 &   1.468  &  -4.894   &  .448
 &  .459   &  .610   &   166 &  1.04 \\
  8.  &     .642 &   1.391  &  -4.779   &  .541
   &  .531   &  .656   & \    95 &    0.97\\
\hline
\end{tabular}
\end{table}
We noted above that the duration and peak flux are independent
stochastic variables. Since the sample was splitted into
subsamples by the peak flux, this means that the parameters of Gaussians
distributions
referring to $T_{90}$ either in the Table 2. or in Table 3.
should be identical within the
statistical uncertainty of estimation. Inspecting $a_x$ in these
Tables - which summarize the results of the EM algorithm - clearly
demonstrates that their difference is much less than $\sigma_x$.
It is also possible
to compare the mean slopes obtained by weighting the results
for the short and long GRBs, respectively,
 in order to test the significance of
the difference between these groups.  One may compute a $\chi^2 =
(m_1-m)^2/\sigma_1^2+(m_2-m)^2/\sigma_2^2$ variable based on the
assumption that the $ m_1$, $ m_2$ slopes of the short and long
GRBs differs from the $m$ weighted mean only by chance. Making
this assumption one obtains  $\chi^2 = 22.2$ indicating that the
null hypothesis, i.e. $m_1=m_2$, should be rejected on a $4.7\sigma$
significance level. {\it The two slopes are different.}

\subsection{Possible sources of the biases}
\label{sec:bias}
 The relationships derived in the previous
Subsection refer to the observed values of GRBs. There is a
dilemma, however, how representative they are for the true
quantities of GRBs not affected by the process of detection. We
mentioned already several major source of bias. Here we summarize
them again:
\begin{itemize}
\item[-] Some GRBs below the threshold remain undetected. Therefore,
the  stochastic  properties of the observed part of the true joint
distribution of $ \left\{\log T_{90},\   \log
F_\mathrm{tot}\right\}$ are not necessarily relevant for the whole
population.
\item[-] Observed duration refers to the detected part of GRBs. The real
 duration might be much longer.
\item[-] There is a similar bias also for the fluence.
\item[-] Additionally, due to the
limited spectral response of BATSE, a significant fraction of the  high energy
 part of  the fluence may  remain unobserved.
\item[-] There is a special bias at short GRBs.
At GRB, where the duration is
shorter than the time resolution of detection, there is a one-to-one
correspondence between the peak flux and the fluence.
\end{itemize}

\subsubsection{Effect of the threshold}

Using the law of full probability we decomposed the observed joint
probability distribution of $ \left\{\log T_{90},\  \log
F_\mathrm{tot}\right\}$ into the distribution of the peak flux and
a conditional probability, assuming $p$ is given. Since the
detection proceeds on three different time scales, one does not
expect a sharp cut on $G(p)$ but the distortion is more
complicated in the reality. Although the observational threshold
may seriously affect the detected form of $G(p)$, it need not
necessarily modify $P(F_\mathrm{tot}, T_{90}|p)$. The detection
threshold, however, may distort also the fluence and duration
themselves, and in this way also the form of $P(F_\mathrm{tot},
T_{90}|p)$.

\subsubsection{True vs. observed duration}
Depending on the light curve of the GRBs a significant fraction of
the outbursts may remain unobserved. So the duration derived from
the observed part is only a lower limit for the true one.
Approaching the detection threshold this effect should become more
and more serious.  Assuming a Gaussian form for
 $P(F_\mathrm{tot}, T_{90}|p)$ one expect a systematic change in the parameters as
one is approaching the threshold. Inspecting the mean values and
standard deviations of the duration in the Tables of the Appendix,
one may really recognize this effect in the three faintest bins.
In the remaining part of the sample, however, there is a
remarkable homogeneity in the mean value and standard deviation of
duration. It is also worth mentioning that the same is true in
Table 2. and 3. summarizing the result of the ML fitting. So one
may conclude that this bias does not play a significant influence
in the 4.--8. bins used for our calculations.

\subsubsection{True vs. observed  fluence}
Similarly to the duration also the observed fluence might be a
lower bound depending on the light curve of the burst. Although
fixing $p$ resulted in a similar functional (Gaussian) form of the
fluence as of the duration, its mean value $a_y$ differs from bin
to bin due to the dependence of $F_\mathrm{tot}$ on the peak flux.
Its standard deviation $\sigma_y$, however, shows a noticeable
homogeneity within the limits of statistical uncertainty. Again,
this implies a constancy in the functional form of
$P(F_\mathrm{tot}, T_{90}|p)$ in the bins studied. The only
exception is perhaps the bin No. 8 for the short GRBs, where the
standard deviation and the $r$ correlations coefficient seems to
depart considerably from the others in Table 2.  One may test the
significance of the excursion of $\sigma_y$ in bin No. 8 by
performing a $F$ test (see e.g. \cite{KS76}). Computing the
$F=\sigma^2_8/\sigma^2_5$ value, where the indexes refer to the
serial number of bins, one obtains $F=2.11$ indicating significant
difference on the 99.9 \% level. Except for this significant
excursion in the 8th bin, the $\sigma_y$ values are statistically
identical implying that the functional relationship between
$F_\mathrm{tot}$ and $T_{90}$ is not significantly influenced by
the process of detection in the bins studied.

\subsubsection{Bias from the spectral response}
\label{spbias} BATSE were observing in four energy channels. Even
the highest energy channel was not able to detect the hardest
parts of the bursts. A significant fraction of the incoming energy
might remain unobserved. In principle, there is a possibility for
estimating the amount of unobserved part of radiation by supposing
a spectral model for the GRB. Fitting this model to the values
measured in the four energy channels one may get an estimate for
the unobserved part. Supposing, the energy distribution of the
bursts can be described by two power laws separated by an
$E_\mathrm{p}$ energy \cite{lp99} did a four parameter fit (two
powers, $E_\mathrm{p}$ and an amplitude) for GRBs detected by
BATSE. A basic trouble at this approach appears in the fact that
numbers of points and parameters to be fitted are identical and,
consequently, any uncertainty in the measured values has a very
sensitive impact on the parameters estimated. Moreover, a
significant fraction of GRBs does not have a reliable fluence in
the high energy channel which exceeds at least the $3\sigma$ level
of the background. In particular, it is true for the No. 4.--8.
bins.

It is well-known that the short bursts are harder in the average
than the long ones. Consequently, the fraction of the unobserved
part of the energy spectrum may have a negative correlation with
the duration in the case of this subgroup. The detected part of
the fluence experiences therefore a positive correlation, assuming
there is no intrinsic relationship between the duration and the
true total fluence. In the case of a real intrinsic relationship
between these quantities, the apparent correlation from the
spectral bias may have a contribution to the real one. One may
expect that the spectral bias is more serious at bursts, where the
whole high energy fluence is buried into the background noise. So
one expect a gradual change in the slope of the relationship
between $F_\mathrm{tot}$, and $T_{90}$ as one proceeds from the
faint bursts to the brighter ones. Table 4. and Table 5. summarize
the frequency of bursts having different $S/N$ ("signal-to-noise")
ratios within the studied peak flux bins, separately.
\begin{table}
  \centering
\caption{Frequency of S/N ratios of fluences in the high energy
channel for GRBs of $T_{90} < 2s$ (the integer numbers given in
the header are the truncated S/N values).
\newline}
\begin{tabular}{|c|rrrr|r|} \hline
          Bin &         &        &   S/N      &         &  Row \\ \cline{2-5}
                  &    .00 &   1.00 &    2.00 &    $>3.00$ & Total \\
\hline
           \  1.  &     0  &     1  &     0   &     0   &     1 \\
           \  2.  &     3  &     0  &     1   &     1   &     5 \\
           \  3.  &    19  &    10  &     7   &     8   &    44 \\
           \  4.  &    57  &    15  &    14   &    18   &   104 \\
           \  5.  &    53  &    28  &    14   &    31   &   126 \\
           \  6.  &    17  &    16  &    17   &    29   &    79 \\
           \  7.  &     4  &     3  &     4   &    36   &    47 \\
           \  8.  &     2  &     4  &     4   &    24   &    34 \\
           \  9.  &     0  &     0  &     3   &    19   &    22 \\
             10.  &     1  &     0  &     1   &     4   &     6 \\
             11.  &     0  &     0  &     0   &     5   &     5 \\
             12.  &     0  &     0  &     0   &     2   &     2 \\
             13.  &     0  &     0  &     0   &     1   &     1 \\
\hline
            Column &   156 &    77 &     65   &   178   &   476 \\
             Total &  32.8 &  16.2 &   13.7   &  37.4   & 100.0\% \\
  \hline
\end{tabular}
\end{table}

\begin{table}
 \centering
  \caption{Frequency of S/N ratios of fluences in the high energy
channel for GRBs of $T_{90} > 2s$ (the integer numbers given in
the header have the same meaning as in Table 4.) \newline}
\begin{tabular}{|c|rrrr|r|} \hline
              Bin &         &         &   S/N   &        &  Row \\
              \cline{2-5}
              &     .00 &    1.00 &    2.00 &    $>3.00$ &Total \\
\hline
          \  1.  &     2  &     1  &     0  &     1  &     4 \\
          \  2.  &    71  &    19  &     7  &    11  &   108 \\
          \  3.  &   193  &    58  &    37  &    53  &   341 \\
          \  4.  &   135  &    65  &    42  &    88  &   330 \\
          \  5.  &    72  &    33  &    40  &    94  &   239 \\
          \  6.  &    33  &    23  &    16  &   103  &   175 \\
          \  7.  &    12  &     9  &    10  &    88  &   119 \\
          \  8.  &     1  &     2  &     7  &    51  &    61 \\
          \  9.  &     1  &     1  &     1  &    49  &    52 \\
            10.  &     0  &     0  &     0  &    33  &    33 \\
            11.  &     0  &     0  &     0  &    14  &    14 \\
            12.  &     0  &     0  &     0  &    13  &    13 \\
            13.  &     0  &     0  &     0  &     5  &     5 \\
            14.  &     0  &     0  &     0  &     2  &     2 \\ \hline
            Column &    520 &   211  &   160  &   605  &  1496 \\
          Total &   34.8 &   14.1 &   10.7 &  40.4  &  100.0 \% \\
  \hline
\end{tabular}
\end{table}
It is clear from Table 5. that the long faint bins are dominated
by bursts with no significant high energy fluence. The contrary is
true for the brighter ones. Proceeding from the faint burst to the
bright ones one does not see a gradual change in the slope of the
$\{ \log F_\mathrm{tot}, \log T_{90}\}$ relationship. Hence, we
may conclude that the spectral bias makes only a marginal
contribution, and the correlation observed is close to the real
one. For the short bursts (Table 4.), in the contrary, a
significant change is observed, which might be interpreted as a
clear sign of spectral bias. It implies, furthermore, that the
real slope, if any, is smaller than the observed one. This fact
strengthens the conclusion on the difference between the short and
long GRBs with respect of the $\{ \log F_\mathrm{tot}, \log
T_{90}\}$ relationship.

\subsubsection{Bias from the finite time resolution}

We mentioned above the detection proceeded on three ($64 \ ms, 256
\ ms$ and $1024 \ ms$) time scales. The incoming photons were
binned in these time scales and the bin having the maximum count
rate were used for triggering the detection. The bursts having
$T_{90} < 64 \ ms$, however, consist of only one bin,
consequently, the fluence and the peak flux are based on the same
incoming photons on this time scale. If the incoming photons of a
burst had the same energy fixing $p$ would mean fixing
$F_\mathrm{tot}$ as well and Eq.~(4) is no longer valid since
$T_{90}$ does not have any impact on the fluence observed. By
fixing $p$ this effect degenerate the distribution of
$F_\mathrm{tot}$ into one point and it does no longer reflects the
distribution of $T_{90}$ we supposed. In the reality, however, the
energies of the incoming photons have a wide range and this effect
is not so pronounced.

As the duration covers an increasing number of bins of $64 \ ms$
the particular  bin representing the peak flux has a decreasing
impact on the value of the fluence. In Table 6  we gave some
stochastic parameters (mean, standard deviation, correlation) of
the joint distribution of $ \log F_\mathrm{tot} $ and $ \log
p_{64} $ within the first 10
 bins of $T_{90}$ of $64 \  ms$,  in order to
see the possible quantitative differences. Except the mean value
of  $ \log F_\mathrm{tot} $ in the first bin, which deviate from
the sample value at about $1\sigma$ level there is no striking
differences between the parameters. For testing the possible
differences between the bins in Table 6  we did a multivariate
analysis of variance (MANOVA) which compares the variances and
covariances of variables within the bins and between them. The
analysis resulted in a difference on the 99.5 \% significance
level. The MANOVA module of the SPSS software package was used for
these calculations \footnote{SPSS is a registered trademark. See
\cite{SPSS} in references}. Repeating the calculation but
abandoning the first bin, the suspected outlier, the significance
dropped back to 50.4 \% inferring  that the distributions in bins
2.--10. were identical within the limits of statistical
uncertainty. Even if we treated the excursion of the bin No. 1 as
a real effect there is only a small number of GRBs in it (see
Table 1) which do not affect the final results in Table 2 and 3.

Summing up the discussions we performed in this subsection on the
different bias we may conclude that either they do not have a
significant impact on the final result (i.e. there is a
significant difference in the $\{ \log F_\mathrm{tot}, \log
T_{90}\}$ correlation between the short and long GRBs) or  the
observed difference in the relationship is even  enhanced in the
reality if  we considered the bias properly.

\begin{table}
\label{t064}
\caption{Mean values and standard deviations of the total fluences
and the 64 ms peak fluxes within the first ten 64 ms bin of the
$T_{90}$ duration. Except the fluence in the first bin all the
values do not differ from those  of the entire sample, within the
limits of statistical uncertainties.  \newline}

\begin{tabular}{|c|cc|cc|c|c|}
\hline & \multicolumn{2}{c|}{ $ \log F_\mathrm{tot} $ } &
\multicolumn{2}{c|}{ $ \log p_{64} $ } &  &
\\ \cline{2-3} \cline{4-5}
   Bin   &    mean    &  st. dev. & mean   &  st. dev.  & corr. coeff. &  no. of GRBs
\\ \hline
 \   1. &  -7.0243   &    .5043  &  .3535    &  .1912  &    .7625    &   \     17 \\
 \   2. &  -6.6280   &    .5109  &  .3815    &  .2447  &    .6333    &   \     33 \\
 \   3. &  -6.6756   &    .5122  &  .3690    &  .2379  &    .6335    &   \     37 \\
 \   4. &  -6.4863   &    .5408  &  .4090    &  .2629  &    .5793    &   \     36 \\
 \   5. &  -6.5480   &    .4428  &  .3691    &  .2460  &    .7125    &   \     26 \\
 \   6. &  -6.5804   &    .5637  &  .3482    &  .2370  &    .6117    &   \     16 \\
 \   7. &  -6.4492   &    .3823  &  .4191    &  .2278  &    .2241    &   \     31 \\
 \   8. &  -6.3312   &    .4756  &  .4278    &  .2802  &    .6868    &   \     20 \\
 \   9. &  -6.4532   &    .4517  &  .3348    &  .2382  &    .6780    &   \     16 \\
    10. &  -6.4292   &    .3826  &  .3682    &  .2258  &    .5620    &   \     16 \\
\hline
 entire & & & & & & \\
 sample &  6.5599   &    .5015  &  .3828    &  .2395  &    .5822    &        248 \\
\hline
\end{tabular}
\end{table}

\section{Discussion}

\label{sec:disc} We have presented evidence indicating that there
is a power-law relationship between the logarithmic fluences and
the logarithmic $T_{90}$ durations of the GRBs in the Current
BATSE Catalog, based on the EM maximum likelihood estimation of
the parameters of the bivariate distribution of these measured
quantities. This relationship holds for both subclasses of GRBs
separately. As shown in the Appendix, the dispersions of the
$T_{90}$ do not differ significantly from those of the $T_{50}$
distributions, and therefore the same correlations and the same
power-law relations would be expected if one used the $T_{50}$
instead of the $T_{90}$. We have also evaluated the possible
impact of instrumental biases, with the results that the
conclusions do not change significantly when these effects are
taken into account.

An intriguing corollary of these results is that the exponents in
the power-law dependence between the fluence and the duration
differs significantly for the two groups of short ($T_{90} < 2$ s)
and long ($T_{90} > 2$ s) bursts at a $4.7\sigma$ level. As shown
in \S \ref{sec:bias}, this also means that the same power law
relations hold between the total energy emitted ($E_\mathrm{tot}$)
and the intrinsic durations ($t_{90}$) of the two groups.The
intrinsic nature of this relation is also confirmed by further
calculations based on a principal component analysis.

While an understanding of such power-law relations in terms of
physical models of GRB would require more elaborate
considerations,  we note that there is substantial evidence
indicating the two classes of bursts are physically different.
First, there is the fact that short burst are harder (\cite{k});
this is confirmed also by the analysis of \cite{muk}. Then, there
is evidence that the spectral break energies of short bursts are
larger than for long bursts (\cite{pac01}). The short bursts have
a different spectral lag vs. luminosity ratios than log bursts
(\cite{no00}). Finally, the number of sub-pulses, and the
soft-to-hard evolution is different depending on the duration
(\cite{gup02}).

The results obtained here are compatible with a simple
interpretation where the bursts involve a wind outflow leading to
internal shocks responsible for the gamma-rays
(\cite{rm94,piran99}), in which the luminosity is approximately
constant over the duration $t$ of the outflow, so that both the
total energy $E_\mathrm{tot}$ and the fluence $F_\mathrm{tot}$ are
$\propto t$. If an external shock were involved, (e.g.
\cite{mr93,piran99}), for a sufficiently short intrinsic duration
(impulsive approximation) there would be a simple relationship
between the observed duration and the total energy, $t \propto
E^{1/3}$, resulting from the self-similar behavior of the
explosion and the time delay of the pulse arrival from over the
width of the blast wave from across the light cone. This
relationship is steeper than the one we deduced for long bursts.

The fluence -- duration relation of GRBs which we have discussed here appears
to be physical, and it is significantly different for the short and the
long bursts. For the short ones, the total energy released
is proportional to the $m=0.81$  power of duration of the gamma ray emission,
 while for the
long ones it is proportional roughly to the of  $m=1.11$  power of
the duration. This may indicate that two different types of
central engines are at work, or perhaps two different types of
progenitor systems are involved. It is often argued that those
bursts for which X-ray, optical and radio afterglows have been
found, all of which belong to the long-duration group, may be due
to the collapse of a massive stellar progenitor (e.g.
\cite{pa98,fry99}). The short bursts, none of which have as of
August 2002 yielded afterglows, may be hypothetically associated
with neutron star mergers (e.g. \cite{fry99}) or perhaps other
systems. While the nature of the progenitors remains so far
indeterminate, our results provide new evidence suggesting an
intrinsic difference between the long and short bursts, which
probably reflects a difference in the physical character of the
energy release process. This result is completely
model-independent, and if confirmed, it would provide a
potentially useful constraint on the types of models used to
describe the two groups of bursts.

\section{Conclusions}
In summary, we have presented quantitative arguments in supporting
two new results, namely that there is a power law relation between
the fluence and duration of GRBs which appears to be physical, and
that this relation is significantly different for the two groups
of short and long bursts. In addition, estimations of the concrete
values of exponents were obtained, two. For the short subgroup one
obtains $m \simeq (0.46 - 1.04) $ with the most probable value
around $m \simeq 0.81$. (In the reality, however, this value could
be much smaller due to a possible strong spectral bias).  For the
long subgroup one obtains
 $m \simeq (0.97 - 1.19) $
with the most probable value around $m \simeq 1.11$.The
  difference is significant on the $4.7\sigma$ level.

 For the short ones, the
total energy released is weakly depending on the duration of the
gamma ray emission, while for the long ones it is proportional
roughly to the duration.  While the nature of the progenitors
remains so far indeterminate, our results provide new evidence
suggesting an intrinsic difference between the long and short
bursts, which probably reflects a difference in the physical
character of the energy release process. This result is completely
model-independent, and if confirmed, it would provide a
potentially useful constraint on the types of models used to
describe the two groups of bursts.

  These results were obtained exclusively
from the statistical studies of BATSE data (using the known
redshifts of the observed afterglows, too) applying only the
mathematical Cram\'er theorem and the law of full probability,
respectively. It is highly surprising that these pure mathematical
theorems allowed to obtain these remarkable results.

\begin{acknowledgements}
We are indebted to Dr. G\'abor Tusn\'ady (R\'enyi Institute for
Mathematics), Dr. Chryssa Kouveliotou and Dr. Michael Briggs (NASA
MSFC) for the useful discussions and critique. We are indebted to
an anonymous referee for the valuable remarks. This research was
supported in part through OTKA grants T024027 (L.G.B.), F029461
(I.H.) and T034549, NASA grant NAG-9192 and NAG-9153, (P.M.), and
Czech Research Grant J13/98: 113200004 (A.M.).
\end{acknowledgements}

\appendix
\section{Comparison of $T_{90}$ and $T_{50}$ statistical
properties}

In order to check, whether there is some influence of the time
dilatation on the distribution of $T_{90}$ or $T_{50}$, we compare
here the basic properties of these two quantities in our sample
for the long and the short bursts, separately. We grouped the
data, using the 256 ms peak flux values, into 0.2 bins in
$P_{256}$, and summarized in Tables \ref{tab:a1}.1 and
\ref{tab:a2}.2 the mean values and the corresponding standard
deviations of the logarithmic durations of GRBs in each peak flux
bin. We stress that this does not include any equalization of the
noise level in the various bins, and is not intended as a test of
the time dilatation hypothesis, but rather as a test of whether
dilatation would have any effect on our results. Inspecting the
durations of long ($T_{90} > 2s$) GRBs summarized in Table
\ref{tab:a1}.1 one sees that, except from the brightest and
faintest bins, there is no significant difference in $\log
T_{90}$. The decrease of the duration in the faintest bin is
probably due to the biasing of the determination, namely, the
fainter parts of the bursts cannot be discriminated against the
background, and therefore the duration obtained is systematically
shorter. There is a remarkable homogeneity and no trend in the
standard deviations of the $\log T_{90}$. In the case of the long
burst $T_{50}$ durations, this quantity shows an increasing trend
towards the bursts of fainter peak flux. The shortening in the
faintest bin is probably also due to selection effects. Similarly
to the $\log T_{90}$ values, the same homogeneity can be observed
in the standard deviations also in case of $\log T_{50}$. The
standard deviations are almost the same in both $\log T_{90}$ and
$\log T_{50}$. One can test whether, within our analysis
methodology and with our sample, there is a significant difference
among the binned $T_{90}$ values, and whether the slight trend  in
the $T_{50}$ significantly differs from  zero. To evaluate the
significance of these data we performed a one way analysis of
variance with the ANOVA program from a standard SPSS package.
The ANOVA compares the
variances within sub-samples of the data (in our case within
bins), with the variances between the sub-samples (bins).

In the case of $\log T_{90}$  the probability that the difference
is accidental is 66\%.  In the case of the $T_{50}$ durations the
same quantities (variances within and between bins) gives a
probability of 98.5\% for being a real difference between bins, or
a probability of 1.5\% that there is no difference between the
bins. This figure gives some significance for the reality of a
trend in the data; however, this value of 0.2 explains less than
1/6 of the variance of $T_{50}$ within one bin. We may conclude
that even in this case the variance is mainly intrinsic.

Inspecting the same data in the case of the short duration bursts
(Table \ref{tab:a2}.2) we come to a similar conclusion, i.e. there
is no sign of trends in the durations of the different bins.
Dropping the two faintest bins, which are definitely affected by
biases, and dropping the poorly populated brightest bins, we
arrive by the analysis of variances with ANOVA to probabilities of
53 \% and 92.1 \% for the difference being purely accidental
between bins in $T_{90}$ and $T_{50}$, respectively.

\begin{table}
\label{tab:a1} \caption{GRBs of long duration ($T_{90} > 2 s$).}
\begin{tabular}{|r|rrrr|r|}
\hline $\log P_{256}$ & $\log T_{90}$& $\log T_{50}$ &
$\sigma_{\log T_{90}}$ & $\sigma_{\log T_{50}}$ & No. of GRBs \\
\hline
    -.50  &   1.24  &    .85 &     .48  &    .47  &    49 \\
    -.30  &   1.42  &   1.00 &     .47  &    .50  &   230 \\
    -.10  &   1.48  &   1.08 &     .49  &    .53  &   309 \\
     .10  &   1.46  &   1.02 &     .51  &    .57  &   272 \\
     .30  &   1.51  &   1.01 &     .52  &    .61  &   194 \\
     .50  &   1.43  &    .94 &     .51  &    .59  &   161 \\
     .70  &   1.45  &    .96 &     .48  &    .56  &   104 \\
     .90  &   1.42  &    .83 &     .54  &    .62  &    56 \\
    1.10  &   1.41  &    .83 &     .50  &    .49  &    44 \\
    1.30  &   1.44  &    .88 &     .50  &    .53  &    34 \\
 $>$1.40  &   1.21  &    .68 &     .41  &    .50  &    29 \\
\hline
\end{tabular}
\end{table}

\begin{table}
\label{tab:a2}
\caption{GRBs of short duration ($T_{90} < 2 s$).}
\begin{tabular}{|r|rrrr|r|}
\hline $\log P_{256}$ & $\log T_{90}$ & $\log T_{50}$ &
$\sigma_{\log T_{90}}$ & $\sigma_{\log T_{50}}$ & No. of GRBs \\
\hline
    -.50  &   -.57  &   -.87 &     .55  &    .60  &     7 \\
    -.30  &   -.65  &  -1.01 &     .53  &    .57  &    43 \\
    -.10  &   -.40  &   -.77 &     .49  &    .51  &   103 \\
     .10  &   -.35  &   -.74 &     .35  &    .32  &   105 \\
     .30  &   -.33  &   -.75 &     .39  &    .41  &    75 \\
     .50  &   -.27  &   -.69 &     .35  &    .36  &    54 \\
     .70  &   -.29  &   -.72 &     .36  &    .34  &    25 \\
     .90  &   -.35  &   -.76 &     .39  &    .36  &    22 \\
    1.10  &   -.18  &   -.72 &     .44  &    .39  &     7 \\
    1.30  &   -.74  &  -1.21 &     .31  &    .43  &     5 \\
 $>$1.40  &   -.72  &   -.90 &     .00  &    .00  &     1 \\
\hline
\end{tabular}
\end{table}

\end{document}